# Tracing the origins of SARS-COV-2 in coronavirus phylogenies

**Erwan Sallard[1], José Halloy[2], Didier Casane[3,4], Etienne Decroly[5*] and Jacques van Helden[6,7**]**

1) École Normale Supérieure de Paris, 45 rue d'Ulm, 75005 Paris, France. ORCID: 0000-0002-2324-3633.

2) Université de Paris, CNRS, LIED UMR 8236, 85 bd Saint-Germain, 75006 Paris, France. ORCID: 0000-0003-1555-2484

3) Université Paris-Saclay, CNRS, IRD, UMR Évolution, Génomes, Comportement et Écologie, 91198, Gif-sur-Yvette, France. ORCID: 0000-0001-5463-1092

4) Université de Paris, UFR Sciences du Vivant, F-75013 Paris, France.

5) Aix-Marseille Univ, CNRS, UMR 7257, AFMB, Case 925, 163 Avenue de Luminy, 13288 Marseille Cedex 09, France. ORCID: 0000-0002-6046-024X

6) CNRS, Institut Français de Bioinformatique, IFB-core, UMS 3601, Evry, France. ORCID: 0000-0002-8799-8584

7) Aix-Marseille Univ, INSERM, lab. Theory and Approaches of Genome Complexity (TAGC), Marseille, France.

* etienne.decroly@univ-amu.fr, **Jacques.van-Helden@univ-amu.fr

The last two authors have equally contributed

## Abstract

**SARS-CoV-2 is a new human coronavirus (CoV), which emerged in China in late 2019 and is responsible for the global COVID-19 pandemic that caused more than 59 million infections and 1.4 million deaths in 11 months. Understanding the origin of this virus is an important issue and it is necessary to determine the mechanisms of its dissemination in order to contain future epidemics. Based on phylogenetic inferences, sequence analysis and structure-function relationships of coronavirus proteins, informed by the knowledge currently available on the virus, we discuss the different scenarios evoked to account for the origin - natural or synthetic - of the virus. The data currently available is not sufficient to firmly assert whether SARS-CoV2 results from a zoonotic emergence or from an accidental escape of a laboratory strain. This question needs to be solved because it has important consequences on the evaluation of risk/benefit balance of our interaction with ecosystems, the intensive breeding of wild and domestic animals, as well as some lab practices and on scientific policy and biosafety regulations. Regardless of its origin, studying the evolution of the molecular**

**mechanisms involved in the emergence of pandemic viruses is essential to develop therapeutic and vaccine strategies and to prevent future zoonoses. This article is a translation and update of a French article published in Médecine/Sciences, Aug/Sept 2020 (doi: 10.1051/medsci/2020123).**

**Twitter message:** A fact-based evaluation of the different hypothesis about the origin of SARS-CoV-2, including a fully reproducible phylogenetic analysis of coronavirus genomic and protein sequences.

## Contents

## 1. Introduction

SARS-CoV-2 is the third human coronavirus (CoV) responsible for severe respiratory syndrome that emerged in the last 20 years, the two previous ones being SARS-CoV in 2002 (Drosten et al. 2003) and MERS-CoV in 2012 (Zaki et al. 2012). SARS-CoV-2, which causes the COVID-19 disease in humans, spread into a pandemic in early 2020. By November 23, 2020, more than 59 million infections had been reported with at least 1.4 million deaths. The etiological agent of COVID-19 was rapidly identified and by 26 January 2020, 10 viral genomes had been sequenced (R. Lu et al. 2020). Sequence comparisons revealed a 99.98% pairwise identity between those genomes, which is characteristic of a recent emergence.

When the first SARS-CoV-2 isolates were sequenced, the closest coronaviruses available in databases were bat-SL-CoVZXC21 and bat-SL-CoVZC45 strains, isolated in 2015 and 2017 from bats in the Zhoushan region of eastern China, and whose genomes showed 88% identity with SARS-CoV-2 (R. Lu et al. 2020). The SARS-CoV-2 genome sequence is more distant from SARS-CoV (79% identity) and MERS-CoV (50% identity), the viruses responsible for the previous human epidemics. Researchers concluded that SARS-CoV-2 is a new infectious agent belonging to the SARS-CoV family, able of human-to-human transmission, and whose animal reservoir is a bat (P. Zhou et al. 2020a; R. Lu et al. 2020).

Based on phylogenetic inferences, sequence analysis and structure-function relationships of coronavirus proteins, informed by the knowledge currently available on the SARS-CoV-2 virus, we present our re-analysis of the available data, and discuss the different scenarios evoked to account for the origin of this coronavirus. Addressing this question is important not only to understand the causes of the pandemic, but also because the actual events at the origin of the virus should be taken into account for decision-making about science policy.

This article is the English translation and update of a French article published in médecines/sciences (DOI 10.1051/medsci/2020123) on July 10 2020. Since our study included a complete re-analysis by ourselves of the genomic and peptidic sequences, this English translation contains an additional section "Material and Methods". We also added the ferret in Figure 4. We also made some minor revisions, added a short conclusion at the end of each paragraph and discussed a few key articles on the subject that were published after our initial publication.

## 2. Evolutionary origin of the new virus

The zoonotic origin of CoVs is well documented. This family of viruses infects more than 500 species of chiropterans (a mammalian order consisting of more than 1200 species of bats) which represent an important reservoir for CoV evolution, allowing the recombination of viral genomes in animals co-infected by different strains (Hu et al. 2017; Luk et al. 2019; Menachery et al. 2015). It is generally accepted that zoonotic transmission of CoVs to humans occurs through an intermediate host species, in which viruses better adapted to human receptors can be selected, thereby facilitating the species barrier crossing (Cui, Li, et Shi 2019). Vectors of zoonotic transmission can be identified by examining the phylogenetic relationships between new viruses and viruses isolated from animal species living in the regions of emergence.

**Figure 1A**, which presents the phylogenetic tree produced from full-genome alignments of different CoVs, shows the close proximity (99% genome identity) between the coronaviruses responsible for the two previous epidemics and the strains isolated from the last intermediate hosts before transmission to humans: civets for SARS-CoV in 2003 (**Figure 1B**)(Guan et al. 2003; Song et al. 2005), and camels for MERS-CoV (**Figure 1C**) (Sabir et al. 2016). In the latter case, several zoonotic transmissions (from animal hosts to humans) have been demonstrated.

Although no epidemic related to direct bat-to-human transmission has been identified to date, experimental studies have shown that more than 60 chiropteran CoVs are capable of infecting cultured human cells (Luis et al. 2013; Menachery et al. 2015). The identification, in 2017, of viral isolates very similar to SARS-CoV in bats raises the issue of a possible direct transmission from chiropterans to humans, which could result from mutations in the receptor-binding domain (RBD) of the viral spike protein having enabled its entry into the host cell (Hu et al. 2017).

In summary, mechanisms of viral emergence and spreading have to be elucidated, and molecular phylogeny can contribute to provide cues about the possible paths of transmissions from bats to humans.

## 3. SARS-CoV-2: from Yunnan to Wuhan?

The origin of SARS-CoV-2 is a matter of debate. Bioinformatic studies revealed that it has a 96.2% identity with a CoV genome (RaTG13) reconstructed from faeces and anal samples of *Rhinolophus affinis* bats. Interestingly, these samples were collected in 2013, but the full genome sequence was only published in early February 2020 (P. Zhou et al. 2020a). Unfortunately, the precise location of the sample collection is documented neither in the original article nor in the sequence databases. However, when the current work was led, we found an exact match between RaTG13 and a 370 nucleotide fragment published in 2016 (KP876546), encoding a BtCoV/4991 polymerase domain, which had been sequenced from isolates collected from a mine shaft in Yunnan Province following the death of 3 miners from an atypical pneumonia (Ge et al. 2016). We thus inferred that this strain resulted from this mineshaft. In the meantime, concerns have been raised by the scientific community about this lack of information concerning the source of the RaTG13 strain, and a recent addendum to the publication confirmed our deduction (P. Zhou et al. 2020b).

More recently, a metagenome (RmYN02) was assembled from faeces samples of 11 bats of the species *Rhinolophus malayanus*, collected in 2019 in Yunnan province. This sequence has 97.2% identity with the first two thirds of the SARS-CoV-2 (ORF 1ab) genome. However, on the remaining third of the genome it diverges quite strongly, especially at the level of the S1 protein and ORF 8 (**Figure 2**) (H. Zhou et al. 2020).

Thus, even though viral strain sequencing enabled the identification of several viruses related to SARS-CoV-2, the genetic distance is still too high to consider them as the proximal ancestors.

## 4. An evolutionary history by fragments

The length of CoV genomes is about 30,000 nucleotides, which is exceptionally long for an RNA virus (by comparison, the length of AIDS and Ebola virus genomes are about 10,000 and 19,000 nucleotides, respectively). CoVs are able to maintain such a long genome thanks to a replication error correction system unique in the world of RNA viruses, that ensures a proofreading mechanism limiting the mutation rate (Eckerle et al. 2010; Ferron et al. 2018; Casane, Policarpo, et Laurenti 2019). The first two-thirds of the genome corresponds to a single gene, ORF1ab, coding for a polyprotein precursor, which is then cleaved into 16 proteins forming the replication/transcription complex. The last third contains 9 genes coding for proteins produced from subgenomic RNAs synthesized by viral polymerase (**Figure 2A**).

The CoV viral polymerase, in addition to its canonical RdRp activity, is able to jump between different RNA strands during replication, a property that probably plays a key role in the recombination capacity of CoVs, and promotes their evolution and host change. Genomic recombination is frequent in chiropteran CoVs (Hu et al. 2017), and is thought to have played a role in the emergence of SARS-CoV in 2002 (Graham et Baric 2010). It is believed that the SARS-CoV genome is a "mosaic" genome composed of pieces of at least two pre-existing CoVs.

A recombinant genome can be detected by comparing its Percent of Identical Positions (PIP) profiles obtained by aligning different genomes to a reference genome. A recombination site is evidenced by the fact that the profiles of two different strains cross each other. In the genomic PIP profiles comparing SARS-CoV-2 with other genetically related viruses (**Figure 2 A-B**), recombinations appear in multiple regions, for example 2,900-3,800, 21,000-24,000, 27,500-28,500 (highlighted with a yellow background).

This mosaicism biases genome-based phylogenies because the inferred tree is a combination of the different evolutionary histories of the recombinant fragments. A phylogenetic inference should thus be made for each recombinant region separately, as illustrated in **Figures 2C** to **2E**, where we have inferred evolutionary trees respectively from the genomic sequences of the ORF1ab polyprotein (**Figure 2C**), the S1 subunit (**Figure 2D**), and the receptor binding domain (RBD) (**Figure 2E**). There are several striking differences between these trees, in particular concerning strain BtYu-RmYN02, which occupies different positions depending on the genomic region considered.

CoV genomes thus contain traces of recombinations and the evolutionary history of coronaviruses should be interpreted in the light of this well-described mosaicism.

## 5. Of bats and men... plus some pangolins?

If we zoom in on the S gene (**Figure 2B**), we can see a decrease in the PIP between the bat strain RaTG13 and SARS-Cov-2 in the RBD-coding genomic region. In particular, at positions 1200 to 1600 of this gene, the PIP falls to 70% while it is higher than 96% over the rest of the genome. In the same region, the closest sequence to SARS-CoV-2 is that of a metagenome (MP789), obtained by assembling pangolin samples (Lam et al. 2020).

Upstream of the RBD, the PIP between SARS-CoV-2 and MP789 is fairly low (60%), whereas downstream it exceeds 90%. This led Xiao and co-workers (Xiao et al. 2020) to hypothesize that SARS-CoV-2 could result from recombinations between viruses infecting bats and pangolins respectively (**Figure 1D**). It should however be noted that, even in Pangolins, there is currently no known non-human virus whose PIP with SARS-CoV-2 exceeds 89% in the RBD region. This level of identity is much lower than the percentages observed between human viruses and strains of the last animal intermediates during previous zoonotic transmissions. For example, the identity rate between the human SARS-CoV genome and that of the closest civet strain is 99.52%.

The initial hypothesis was thus that SARS-CoV-2 results from multiple recombinations between different bat and pangolin CoVs, followed by an adaptation that would have increased its capacity for human-to-human transmission. Transmission to humans would come from contact with the intermediate host sold on the Wuhan market (Liu et al. 2020). However, this hypothesis raises many questions. On the one hand, the first identified patients did not attend the Wuhan market (Huang et al. 2020). On the other hand, despite the search for viruses in the animal species sold on this market (Zhang, Wu, et Zhang 2020), to date no intermediate virus has been identified that may result from the supposed recombination between a bat virus and a pangolin virus. Moreover, the source of the pangolin samples is still unclear, and a warning has recently been posted on Nature website (on Nov 11, 2020) about Xiao's article (Xiao et al. 2020), indicating "additional actions will be taken once this matter is resolved".

Until the last hypothetical recombinant has been identified and its genome sequenced, it will not be known for certain in which species this recombination has taken place: a bat, a pangolin, another species? And above all, in which conditions? It is conceivable that the recombination took place in farm or laboratory animals rather than in pangolin or wild bats: in the former case, transmission to humans would be favoured by closer and more frequent contact. Furthermore, the human ACE2 (angiotensin converting enzyme 2) protein, which is used by SARS-CoV-2 as a receptor for cell infection, is closer to the homologous protein of numerous farm animals than to the ACE2 proteins of pangolins and bats (**Figure 4**). Another hypothesis is that the similarity between the RBD sequences of pangolin and SARS-CoV-2 results from a convergent evolution.

The hypothesis promoted by most specialists is that the virus has a zoonotic origin. This hypothesis relies on phylogenetic studies suggesting two main scenarios to explain the origin of SARS-CoV-2: (i) adaptation in an animal host before zoonotic transfer, or (ii) adaptation in humans after zoonotic transfer (Latinne et al. 2020; P. Zhou et al. 2020a; Lam et al. 2020; Zhang, Wu, et Zhang 2020; Xiao et al. 2020; Andersen et al. 2020). However, in the absence of evidence regarding the last animal intermediate before human contamination (the "proximal" origin of the virus), some authors suggested that SARS-CoV-2 may have been manufactured in a laboratory (synthetic origin) (Segreto et Deigin 2020). Others suggested that SARS-CoV-2 may result from a chiropteran virus that became adapted to other species in laboratory animal models and then escaped from the laboratory (Sirotkin et Sirotkin 2020). It might also be envisaged that it comes from a viral strain cultured on human cells in a laboratory in order to study its infectious potential, and that has been progressively "humanized" (adapted to humans) by selection of the viruses having the highest ability to spread in these conditions.

Regardless of the mechanism of appearance of the virus, it is important to understand how it crossed the species barrier and became highly transmissible from human to human, in order to protect against new outbreaks (Cheng et al. 2007). In conclusion, there is little evidence that supports pangolins as the intermediate host in a zoonosis and in absence of such evidence additional virus strains should be collected from wild sites and from animal farms in order to elucidate the transmission path from bat to Human.

## 6. Protein S is a major player in the evolution of CoVs and the crossing of the species barrier

The S gene codes for the spike protein, which is located in the viral envelope and forms characteristic crown-like protrusions on the viral surface, from which the name of coronavirus derives (**Figure 3A**). The spike protein plays a decisive role in the initiation of the viral cycle because it participates in the recognition of the ACE2 receptors of the host cell, which then allows the viral genome penetrating the cells. This receptor, present in all mammal species, is located at the outer surface of different human cell types, including alveolar cells in the lung, enterocytes in the small intestine, arterial and venous endothelial cells, and arterial smooth muscle cells in most organs. ACE2 messenger RNA is also detected in the cerebral cortex, striatum, hypothalamus and brainstem. Interferons, which are signaling proteins produced in response to viral infections, also increase ACE2 expression, thereby promoting the systemic spread of the virus (Ziegler et al. 2020).

The S protein is synthesized as an inactive precursor, which requires two successive proteolytic cleavages to ensure its biological function (**Figure 3B**). The first cleavage, called *priming*, generates the S1 and S2 subunits. The second cleavage occurs within S2 and releases the end of a *fusion peptide* located at the beginning of the S2' subunit. These two proteolytic cleavages are likely catalysed respectively by furin and other proteases such as TMPRSS2 (transmembrane protease 2) (Hoffmann et al. 2020). The S protein cleavage is essential for the formation of infectious viral particles, as they allow fusion between the viral and cell membranes (**Figure 3A**).

The S1 protein of SARS-CoV and SARS-CoV-2 contains the RBD domain (**Figure 3B** and **4**) which ensures the recognition of the ACE2 receptor by the virus (Wrapp et al. 2020; Wu et al. 2020; Lam et al. 2020). It also bears most of the exposed sites on the virus surface (**Figure 3C**), including the major antigens that can be recognized by antibodies produced by infected hosts (Ni et al. 2020). The sequence of these exposed sites shows a high variability between virus species, which results from the selection of mutations enabling viruses to escape the immune response.

The RBD residues directly involved in the recognition of ACE2 are also subject to strong evolutionary constraints (**Figure 4**). Some key residues are required for efficient infection of chiropterans, the intermediate host, or humans (G. Lu, Wang, et Gao 2015; Letko, Marzi, et Munster 2020; Yan et al. 2020), and SARS-CoV-2 may have acquired its epidemic propensity through mutations in these key residues. Phylogenetic analysis of the S protein is therefore particularly informative to understand the evolution of CoVs and their ability to cross the species barrier.

In this context, the identification of new SARS-CoV-2-like coronavirus sequences isolated from Malayan pangolins was an important step forward. Indeed, although the whole genome PIP between these viruses and SARS-CoV-2 does not exceed 89% (**Figure 2A**, strain MP789) vs. 96% for RaTG13, the amino acid identity is 98% in the RBD domain (Xiao et al. 2020). This difference of similarity between nucleic acids and proteins can be explained by the fact that almost all mutations in this region are synonymous, suggesting a strong selective pressure, presumably related to the key function of RBD in the infection. It therefore appears that some CoVs that infect pangolins possess an RBD domain very close to SARS-CoV-2, and may thus have a strong affinity for the human ACE2 receptor and thereby infect human cells more effectively than bat viruses (**Figure 4A**).

With the currently available sequences, analyses based on phylogenies of complete virus genomes are not sufficient to draw firm conclusions on the evolutionary origin of SARS-CoV-2. This leads to various alternative hypotheses about a possible synthetic origin of this virus: for example, it was proposed that SARS-CoV-2 was reconstructed from metagenomic sequences obtained from bat faecal samples. Concerns have also been raised in relation to genetic manipulations of viruses in order to understand the mechanisms driving the crossing of species barriers. Indeed, experiments of serial passage between model animals and/or cultured cells would lead to a fast evolution by selecting adaptive traits resulting from spontaneous mutations (Sirotkin et Sirotkin 2020).

## 7. Genetic manipulations of viruses and gain of function experiments

The issue of the natural or synthetic origin of SARS-CoV-2 deserves to be examined in more detail on the basis of available evidence. Hypotheses must be examined knowing which type of genetic manipulations are currently carried out in laboratories. Indeed, the manipulation of genomes of potentially pathogenic viruses is a common practice, which aims at understanding the mechanisms of replication and emergence of these viruses, and at developing new antiviral or vaccine strategies. Due to the risks of unexpected species crossing, contamination of a new host (especially humans) and accidental dissemination of artificial recombinant viruses, these investigations are conducted in high-security laboratories (BSL3 or BSL4) subject to strict control and transparency procedures.

The controversy on gain of function experiments (increase in virulence or infectivity of viruses by genetic manipulation) began in 2011, following the work of the teams of Ron Fouchier (Russell et al. 2012) and Yoshihiro Kawaoka on the influenza virus (Imai et al. 2012). In order to understand the virulence factors of influenza, these researchers had tested the effect of mutations that could increase the transmissibility of the H5N1 virus in different animal models. The U.S. Department of Health's National Science Advisory Board for Biosecurity (NSABB), alerted by these experiments in December 2011, asked the journals Nature and Science not to disclose the results of this work on behalf of the significant death toll expected in case of intentional (bioterrorism) or accidental release of these viruses from the laboratory. Because of the importance of the results for public health and the research communities, the NSABB ultimately recommended the general findings to be published, but recommended that the manuscripts should not include "methodological and other details that might allow replication of the experiments by those who would seek to do harm" (Committee on Science et al. 2013).

The risk of accidental escape of new potentially pandemic pathogens are increased by the proliferation of high biosafety laboratories (BSL-3 and BSL-4) in densely populated areas (Van Boeckel et al. 2013). In addition, experiments on viruses such as avian influenza viruses or SARS from chiropterans, that are currently unable to infect humans, are allowed in BSL-3 laboratories: it increases the risk of accidents because selection or mutagenesis can confer an epidemic potential to these viruses (Enserink 2003; Normile 2004; Henkel, Miller, et Weyant 2012).

Prior to 2002, although they caused major epidemics in livestock, CoVs were considered to be viruses of low public health significance, as they were mainly responsible for benign diseases such as seasonal colds. Since the emergence of SARS-CoV in 2002, studies conducted in the United States and in China have tested the possibility of zoonotic transfer of bat CoVs to humans and attempted to elucidate the processes leading to the emergence of new pathogens (Ren et al. 2008; Zeng et al. 2016; Menachery et al. 2015; Hu et al. 2017).

Recombinant viruses potentially adapted to humans have been constructed from bat CoVs, including through replacement of the bat RBD with the RBD of human SARS-CoV in US and Chinese laboratories (Zeng et al. 2016; Menachery et al. 2015). Among other discoveries, these experiments nevertheless revealed that infection of human cells is often limited because the activation of the S protein requires specific proteolysis, which is incompletely performed by human cells (**Figure 3B**). This difficulty can be circumvented by treating viruses with trypsin (Menachery et al. 2020) or by adding a furin proteolysis site downstream of the RBD domain, which can be cleaved by human cells (Follis, York, et Nunberg 2006; Belouzard, Chu, et Whittaker 2009). These investigations indicate as expected that it is possible to adapt bat viruses to infect human cells or various animal models, and that chiropteran CoVs have the potential for direct zoonotic transmission to humans, particularly if they acquire an adapted proteolysis site, which requires only a few mutations or the insertion of a short sequence rich in basic amino acids (Hu et al. 2017). This hypothesis has been put forward by Sirotkin et al. who developed the hypothesis that the virus might have arisen from serial passages, and accidental escape from the lab (Sirotkin et Sirotkin 2020).

The spectacular progress in synthetic biology and reverse genetics methods over the last 20 years also increases the risks associated with gain of function experiments: it is now possible to assemble a viral genome in about ten days from different DNA fragments synthesized on the basis of sequences from one or more wild virus genomes and to obtain a "new" virus in less than a month (Zeng et al. 2016; Thao et al. 2020; Iseni et Tournier 2020).

# 8. HIV sequences and a furin cleavage site inserted into the SARS-CoV-2 S gene?

Doubts about the zoonotic origin of SARS-CoV-2 were raised following the observation of 4 insertions of short sequences (noted i1 to i4 in **Figures 3C**, **2B** and **5A-D**) within the S1 protein. The fourth insertion (i4) is particularly noteworthy, because it is unique among all the coronaviruses of the SARS group, and because it confers a particular property to the protein (Coutard et al. 2020)). This insertion adds 4 amino acids at the precise cleavage site between S1 and S2, immediately upstream of an arginine (**Figure 5D**), which creates a sequence RRAR, corresponding to the consensus recognition motif of the furin protease. Similar changes in the cleavage site of viral envelope proteins are known to promote infectivity of different respiratory viruses (e.g. Influenza or Sendaï), by facilitating their spread through the respiratory tract and systemic dissemination (Moulard et Decroly 2000; Sun et al. 2010).

The uniqueness of this furin cleavage site in the spike protein of SARS-CoV-2, as well as its conservation in all the isolates of SARS-CoV-2 circulating in human populations suggest that it has favoured, if not allowed, the crossing of the species barrier and/or the evolution of the epidemic form of the virus. The importance of this conservation for human-to-human transmission is supported by two further observations: first, this proteolysis site is unstable when the virus is grown on some cultured simian cells (VeroE6 strain), and second, experiments on hamsters show reduced symptom severity when the furin site is deleted (Lau et al. 2020). This suggests that a strong selection pressure is exerted on this furin site for the spread of SARS-CoV-2 in humans.

It should also be noted that the appearance of furin cleavage sites in human CoVs is not an exceptional event. Similar sites have been observed in human CoVs outside the SARS-CoV group, such as MERS, HKU1, OC43 (Matsuyama et al. 2018; Coutard et al. 2020). Such insertion could result from the presence of palindromic sequences found around the furin cleavage site, thereby providing a natural mechanism to explain the insertion of proteolytic cleavage site (Gallaher 2020).

Three other insertions were identified (**Figure 5A-C**): these short sequences are present in SARS-CoV-2 but absent from some chiropteran isolates (i.e.: CoVZC45 and CoVZXC21) and from SARS-CoV. The authors of a prepublication (Pradhan et al. 2020) pointed out a fact they qualify uncanny: at these four insertions, the SARS-CoV-2 S protein shows similarities with fragments of the HIV-1 virus ENV and GAG proteins. However, following critical comments regarding methodological and interpretation weaknesses, the authors withdrew their manuscript from the bioRxiv site.

This "uncanny fact" should therefore have remained anecdotical. Nevertheless, in April 2020, Professor Luc Montagnier, recipient of the 2008 Medicine Nobel Prize for his contribution to the discovery of HIV, made the headlines by claiming on several media that these insertions could not result from natural recombination or accident, but of man-made genetic manipulations, carried out intentionally, presumably as part of a research aimed at developing HIV vaccines. These assertions were immediately challenged by numerous scientists, who argued that the similar sequences between HIV and SARS-CoV-2 are so short (about 30 nucleotides in a genome of 30,000) that their similarity is likely coincidental. The controversy further amplified, in a politically tense context where the President of the United States accused China of having let the virus escape from a BSL-4 laboratory in Wuhan.

Such a controversial climate does not favour a rational analysis of the facts and, paradoxically, no in-depth analysis has been published to date on the origin of these insertions. Yet, as we show below, bioinformatics and molecular phylogeny approaches can provide interesting new information.

Luc Montagnier's hypothesis is based on an analysis of sequence similarities between a fragment of the SARS-CoV-2 S gene and the HIV genome. The most significant alignment obtained by replicating this analysis is shown in **Figure 6A**. The significance of the alignment is reflected by BLAST *expect* score, which estimates the statistical expectation, i.e., the number of matches of the same level of similarity that would be found if random sequences were used as queries. A similarity between two sequences is considered significant when the "*expect* score" is much lower than 1. For example, when comparing homologous gene sequences, scores in the order of $10^{-150}$ are frequently found. On the other hand, an *expect* score higher than 1 – such as observed here – means that the similarity is insufficient to support a common ancestral origin of the sequences. Hence, with an *expect* of 7.5, the alignment of HIV and SARS-CoV-2 sequences does not indicate any sign of homology. This can easily be tested by running the same query with a randomised sequence obtained by shuffling the residues of the spike protein. **Figure 6B** shows the result of this test: the random sequence returns matches as significant as the coronavirus gene (**Figure 6A**). This confirms that the similarities between coronavirus and HIV are not significant.

In addition, phylogenetic inferences carried out in the vicinity of the insertions (**Figure 5**) show that the four insertions found in SARS-CoV-2 cover different sub-groups of coronavirus strains, suggesting that they occurred independently at different times of coronavirus diversification. In particular, the first three insertions are observed in virus sequences isolated not only from human and bats (RaTG13), but also from pangolins from China or Malaysia. The hypothesis that these insertions are the result of recent experimental manipulations would not explain the presence of these sequences in several virus isolates from different species, collected at different locations, especially since these insertions occurred at different times during the evolution of these virus strains.

In this context, how can we understand the appearance and function of these insertions? The analysis of S protein alignments shows that insertions occur very frequently in the coronavirus S protein. Moreover, its structure, resolved by electron cryo-microscopy (Wrapp et al. 2020), indicates that the four SARS-CoV-2 insertions are located on its surface (**Figure 3C**), suggesting that they participate in the escape of the virus to the infection control by antiviral immunity.

In conclusion, the presence of HIV inserts inside the spike protein can be ruled out. The phylogeny indicates that insertions/deletions occur frequently in CoV spike, and that the furin cleavage site, which is unique to SARS-CoV-2 among SARS-like coronaviruses, is the most recent insertion. This insertion plays a key role in SARS-CoV-2 spreading, but unfortunately, the data currently available does not enable us to conclude when, where and how this insertion appeared. Identifying the proximal animal host before the zoonosis might bring an answer to this question.

## 9. Current status: the jury is still out

A puzzling question is the origin of the specific features of RBD. It is clear that this RBD cannot come from the 2002 SARS-CoV virus, since SARS-CoV's RBD is genetically very distant from SARS-CoV-2, as shown by PIP profiles as well as the phylogenetic analysis of CoV RBD domains (**Figure 2B, E**). In addition, the SARS-CoV-2 residues that play a key role in the recognition of the ACE2 receptor are not conserved in the SARS-CoV (**Figure 4**). These sequence differences entail a 20-fold higher affinity for the receptor in SARS-CoV-2 than in SARS-CoV (Walls et al. 2020). However, SARS-CoV-2 binding to target cells is comparable to that of SARS-CoV, as the accessibility of the RBD is suboptimal in the former (Shang et al. 2020).

The article "The proximal origin of Sars-coV-2" (Andersen et al. 2020) is recurrently put forward as a proof for the natural origin of SARS-CoV-2. However, their reasoning does not rely on an actual positive proof of the zoonotic origin. Indeed, as discussed above, the divergence between RaTG13 and SARS-CoV-2 dates from several decades and we still don't know any suitable candidates for proximal animal hosts and viral strains. Rather, the rationale of this article consists in opposing two mutually exclusive options, which are implicitly considered as exhaustive: either a "natural proximal origin" (i.e. a recent zoonosis), or a virus intentionally designed on the basis of prior knowledge, and constructed by reverse engineering (design hypothesis). They provide two arguments against the design hypothesis (the prior knowledge was insufficient to conceive the RBD, and there is no trace of reverse engineering in the genome), and thus conclude that the virus must be of natural origin. This reasoning is however flawed, because it restricts the choice to a dichotomy whereas several other hypotheses are conceivable. In particular, the authors do not consider the possibility that the virus would result from laboratory selection through successive passages between animal species or cells, as discussed in detail by Sirotkin and Sirotkin (Sirotkin et Sirotkin 2020). Moreover, even though it is not obvious to identify a posteriori any tracer of genetic manipulation and as discussed above, there are currently several traceless options for genetic engineering. In conclusion, the arguments supporting the natural proximal origin are so far inconclusive and, albeit this hypothesis has been widely supported by the scientific community (Calisher et al. 2020), alternative hypotheses about a possible laboratory origin cannot be formally ruled out. This question should thus be re-opened and all the hypotheses should be evaluated, and weighted according to the different elements of information at our disposal.

## 10. Discussion and perspectives

We have shown above that bioinformatics analysis can shed light on the possible origins of SARS-CoV-2, the virus responsible for the COVID-19 pandemic. This article reports only preliminary analyses, and further studies are currently being carried out in laboratories to dig into the available data and extract all relevant information. It is hoped that new data will soon be available that will resolve the remaining unanswered questions. The current understanding is therefore incomplete and provisional, but it is useful to ask which conclusions can already be drawn on the basis of available data, and what kind of new results or analyses would provide us with additional information, or even enable us to ascertain the origins of the virus.

The first question is that of the last animal host before man. Phylogenetic analyses indicate that CoVs from chiropterans frequently circulate between different bat species and are occasionally transmitted to other mammals. Virus coevolution with their host and adaptation to new hosts involve point mutations but also recombinations, which are frequent in coronaviruses. These raise particular difficulties because whole genome-based phylogenetic inference is biased by the mosaicism, since the resulting tree would reflect a mixture of the distinct evolutionary trajectories followed by the different genomic fragments. It is therefore important to identify the recombinant fragments, and to perform separate phylogenetic inferences for each one. Available data suggest that SARS-CoV-2 is derived from multiple recombination events between chiropteran CoVs that undoubtedly represent the primary reservoir of the virus. The effect of recombinations is particularly important for the adaptability of the S protein because of its key role in the interaction with the host ACE2 protein.

The possible role of pangolin viruses in this process remains uncertain because, although its functional importance is established, the region of strong similarity between pangolin virus and SARS-CoV-2 is short and the likelihood of pangolin-to-human transmission could be very low. Furthermore, even the pangolin viruses that are the closest to SARS-CoV-2 (such as MP789), as well as its bat-CoV relatives (notably RaTG13 and RmYN02) display a relatively low identity rate with SARS-CoV-2, suggesting that closer relatives and potentially more recent intermediate hosts remain to be discovered. The discovery of animal viruses sharing a very high similarity with SARS-CoV-2 would validate its natural origin. Consequently, the sequencing of new CoV genomes potentially involved in zoonosis (those circulating in chiropterans and in species in contact with human populations) is requested. It would be necessary to focus primarily on mammalian species whose ACE2 receptor better matches the key characteristics of the human receptor than chiroptera, such as pigs, goats, sheep, cows or cats (**Figure 4B**). It should be kept in mind that the reliability of the results depends on the quality of sequencing, metagenomic reconstructions, public accessibility of the data and the accuracy of the annotations in sequence databases (Hassanin 2020).

The insertion between the S1 and S2 subunits of the S protein created a furin-sensitive proteolytic cleavage site which appears to contribute to its infectivity and/or epidemic propensity in humans. This insertion must be recent since it is absent from all the close relatives of SARS-CoV-2. This observation is crucial as this site probably played a key role in the species barrier crossing and/or in the efficiency of human-to-human transmission, which is a prerequisite for the emergence of epidemics.

By bringing people into contact with wildlife in nature or in farms, wildlife trafficking and deforestation should also be questioned. In China, pangolin farms have been spreading, raising new health issues, beyond the questions about the feasibility of such domestication (Hua et al. 2015). In addition, these new exotic farms come alongside all intensive farming of domestic animals (poultry, pigs, etc.), which also creates reservoirs of viruses (influenza, etc.) in areas with high human density (Gibbs, Armstrong, et Downie 2009).

Knowing that several laboratories are conducting and publishing gain of function experiments to characterize the interactions between coronavirus RBD and transmembrane receptors such as ACE2, it has been suggested that SARS-CoV-2 would result from experiments to "humanize" an animal virus of the RaTG13 type. To date, no convincing evidence has been reported from the initial studies carried out by the scientific community. However, bioinformatic analyses revealed biases in codon usages that might reflect some genetic manipulation (Gu et al. 2020). Segreto and Deigin develop the hypothesis of a genome modified by molecular engineering (Segreto et Deigin 2020). More thorough analyses are warranted to clarify this issue.

Beyond the frame of existing national regulations (e.g. for France the Microorganisms and Toxins Regulations, MOT), at the global level, the identification, the isolation and the culture of these new respiratory viruses must be carried out under the safest possible experimental conditions, with unquestionable traceability, in order to prevent zoonotic transmission. Considering the impact of infectious risks, civil society and the scientific community will have to re-examine the practice of gain of function experiments and adaptation to humans in the laboratory, of viral strains cultured in intermediate animal hosts. In 2015, aware of this problem, the US federal agencies froze funding for any new study involving these experiments (« Statement on Funding Pause on Certain Types of Gain-of-Function Research » 2015). This moratorium ended in 2017 (Burki 2018). A new assessment of risks versus potential benefits of these practices should be done. Of course, it is desirable to avoid the pitfall of overly strict regulations that would impede the study of the molecular mechanisms involved in the spread of viruses and thereby prevent the development of antivirals and vaccines. Regardless of its origin, the study of the molecular mechanisms involved in the emergence of potentially pandemic viruses is and will remain essential to develop therapeutic and vaccine strategies.

To conclude, on the basis of currently available data it is not possible to determine whether the emergence of SARS-CoV-2 is the result of a zoonosis from a wild viral strain or an accidental escape of experimental strains. Furthermore, a recent zoonosis would justify enforcing the sampling in natural ecosystems and/or farms and breeding facilities in order to prevent new spillover. Conversely, the perspective of a laboratory escape would call for an in-depth revision of the risk/benefit balance of some laboratory practices, as well as an enforcement of biosafety regulations.

## Material and methods

### *Reproducibility of the analyses*

All the analyses to produce the results and figures of this article follow the FAIR principles (Findable, Accessible, Interoperable and Reusable). The software environment, sequence data, commented code and examples can be downloaded as a github repository (https://github.com/jvanheld/SARS-CoV-2_origins), and the main results can be browsed on the github web pages (https://jvanheld.github.io/SARS-CoV-2_origins/). The software environment is fully described in a yaml-formatted conda configuration file, enabling to re-run all the analyses on Linux or Mac OS X operating systems. The release of the code corresponding to this article is available on Zenodo (https://zenodo.org/record/3931505).

A few sequences could however not be made available in the github repository because they were downloaded from the GISAID server, which does not allow to redistribute the sequences. Since these sequences were crucial to reproduce some key elements of the current debate about SARS-CoV-2 origins, we incorporated them in our analyses. These sequences can be found on the GISAID server (https://www.gisaid.org/) with the IDs EPI_ISL_412977 (Bat virus metagenome RmYN02), EPI_ISL_410544 (Pangolin virus Gu-P2S_2019) and EPI_ISL_410721 (Pangolin virus genome Gu1_2019), respectively.

### *Collections of viral strains*

We defined two collections of viral strains enabling us to highlight different aspects of SARS-CoV-2 origins: (1) "around-CoV-2" regroups human SARS-CoV-2 with 18 other strains from Bat or Pangolins that are closer to SARS-CoV-2 thant to any other coronavirus genome; (2) "selected" includes the latter collection plus 23 additional strains representative of other coronavirus groups, including SARS-CoV, MERS-CoV and a few more distant strains.

### *Sequences*

Viral sequence genomes were collected from the NCBI Nucleotide database (https://www.ncbi.nlm.nih.gov/nuccore/). A workbook with the identifiers and descriptions of the sequences is included in the github and zenodo releases. S gene sequences were extracted based on the annotation of their coordinates in the NCBI annotations. S protein sequences were obtained by translating the corresponding gene sequences.

### *PIP profiles*

Profiles of Percent Identical Proteins were computed with an original R script available on the github repository, which enables to draw PIP profiles for either nucleic or peptidic sequences.

For genomic PIP profiles, each viral sequence of interest ("around-cov-2" or "selected" collections) was aligned onto the reference genome (SARS-CoV-2) using the Needleman-Wunsch global pairwise alignment algorithm. PIP profiles for coding sequences were based on translation-based multiple alignments of the nucleic sequences with the R function DECIPHER::AlignTranslation(). The PIP was measured on the resulting aligned nucleic sequences, as well as on the aligned protein sequences (not shown in this article).

### *Phylogenetic analysis*

For nucleotide as well as amino acid sequences, we performed multiple alignments with clustalw v2.1 (Larkin et al. 2007) followed by maximum likelihood-based phylogenetic inferences with PhyML v3.3.20190909 (Guindon et al. 2010). We assumed a GTR substitution model for nucleotide sequences and an LG substitution model for amino acid sequences, with gamma-distributed substitution rates. The other PhyML parameters were left to their default value.

### *Structural analyses of the spike protein*

A model of the full SARS-CoV-2 spike protein was built by aligning the sequence of SARS-CoV-2 spike protein on the PDB 6acc.1.A model of the full SARS-CoV spike protein (the most complete model of a mature coronavirus spike trimmer available to date) using the SWISS-MODEL online tool.

Structural analyses were conducted using the pymol software and the scripts available on https://github.com/jvanheld/SARS-CoV-2_origins/tree/master/scripts/pymol. The 6m0j model and the model we built were aligned and the insertions identified by running the script https://jvanheld.github.io/SARS-CoV-2_origins/scripts/python/detection_insertion.py on the alignment of all sarbecovirus spike sequences (https://jvanheld.github.io/SARS-CoV-2_origins/results/spike_protein/muscle_alignments/selected_coronavirus_spike_proteins_aligned_muscle.clw) with the SARS-CoV-2 spike as the reference sequence were colored depending on the number of coronaviruses in which this insertion is found.

## Supplementary files

### *Table 1. Metadata about the viral genomes and S proteins used in this article.*

File: selected_coronavirus_genomes-and-S-proteins_2020-07-06.xlsx

## Acknowledgements

We would like to thank the following colleagues for their careful revision of the early drafts of the French manuscript, and their many suggestions for its improvement: Cathy Bellan, Mathias Bonal, Bruno Canard, Bruno Coutard, Hélène Chiapello, Denis Gerlier, Catherine Nguyen, Nadia Rabah, Annick Stevens, Denis Thieffry.

## Tables

Table 1. Sources and publication dates for the viral strains discussed in this article.

| Strain | Host | Isolate origin | Isolate date | Publication date | Precisions concerning the origin of the sample |
|---|---|---|---|---|---|
| BtBM48-31 | Bat | Bulgaria | 2008 | Oct 1, 2010 | |
| BtGX2013 | Bat | China | 2013 | Jul 7, 2017 | |
| BtHKU3-12 | Bat | China (unspecified) | unspecified | Apr 5, 2010 | China according to publication but origin not indicated in NCBI |
| BtRaTG13_2013_Yunnan | Bat | Yunnan | Jul 24, 2013 | Mar 24, 2020 | Sequence published in 2020, annotated as isolated in 2013. Partial genomic sequences (RoRp region) published by Shi group of 2016 i have 100% identity with RaTG13 |
| BtRs4874 | Bat | China | Jul 21, 2013 | Dec 18, 2017 | Shi's group in Wuhan (Hubei Province, China) |
| BtYN2013 | Bat | China | 2013 | Jul 7, 2017 | |
| BtYN2018B | Bat | China | Sep 1, 2016 | Jun 30, 2019 | |
| BtYu-RmYN02_2019 | Bat | China, Yunnan - Xishuangbanna | Jun 25, 2019 | Feb 3, 2020 | Metagenome constructed by sequencing a mixture of 11 faecal samples from Rhinolophus malayanus bats |
| BtZC45 | Bat | Zhoushan | 2017 | | |
| BtZXC21 | Bat | Zhoushan | 2015 | Feb 5, 2020 | |
| Cv007-2004 | Civet | China : Guangzhou in Guangdong Province | 2019 | Dec 1, 2005 | Civet virus closest to the 2003 SARS-CoV. Quoted from the article: "These cases were not linked to any laboratory accident." |
| HuCoV2_WH01_2019 | Human | China, Hubei, Wuhan | Dec 23, 2019 | Feb 11, 2020 | Pandemic reference genome for COVID-19 |
| HuSARS-Frankfurt-1_2003 | Human | Frankfurt | 2003 | Mar 16, 2004 | Reference genome for the 2003 SARS epidemic |
| PnGu-P2S_2019 | Pangolin | China, Guangdong | 2019 | Feb 17, 2020 | Sequence available in GISAID, very close to MP789. Pre-publication version ? |
| PnMP789 | Pangolin | China : smuggled Malayan pangolins, Guangdong customs | Mar 29, 2019 | Apr 23, 2020 | Metagenome assembled from samples of 3 pangolins collected in March and July 2019. |
| PnGu1_2019 | Pangolin | China, Guangdong | 2019 | Feb 18, 2020 | |
| PnGX-P1E_2017 | Pangolin | Chinese customs on a flight from Malaysia | 2017 | Apr 23, 2020 | |
| PnGX-P2V_2018 | Pangolin | Chinese customs on a flight from Malaysia | 2018 | Apr 23, 2020 | Collected from pangolin, this strain has been cultured on human cells (and therefore presumably suitable for human infection). |

# Figure legends

**Figure 1. Phylogeny and emergence of coronaviruses. (A)** Tree inferred from complete coronavirus genomes, based on progressive multiple alignment (clustalw) followed by maximum likelihood inference (PhyML). Genomes assembled from metagenomic data are marked with a star. The prefixes of virus names indicate the host species: Bt (Bat), Hu (Human), Pn (Pangolin), Cv (Civet), Cm (Camel), Pi (Pig). Note that the distances between HuCoV2 and the closest viral strains (BrYuRmYN02, BtRaTG13) are higher than for SARS-CoV (Human - Civet) or MERS-CoV(Human - Camel) **(B-D)** Hypotheses of transmission from the animal reservoir (bats) to humans, based on the molecular phylogeny of viral genomes. **(B)** For the SARS-Cov pandemic of 2003, the intermediate host is the civet. Direct bat-to-human transmission is also under consideration. **(C)** Pandemic MERS-CoV of 2012, with the camel as an intermediate host. Several direct transmission events have been documented. **(D)** Pandemic COVID-19. Several scenarios are proposed about the last host before transmission to humans. Distances between HuCoV2 and the closest viral strains are found to be greater than for SARS-CoV (human-civet) or MERS-CoV (human-camel).

**Figure 2. Profiles of Percent Identical Positions (PIP) between SARS-CoV-2 and other coronavirus genomic sequences. (A)** Genome-wide PIP profile (with sliding windows of 800 base pairs). **(B)** PIP profile along the S gene (200bp sliding windows). **(C-E)** Impact of recombinations on the topology of phylogenetic trees inferred from different genomic regions: ORF1ab **(C)**, S1 **(D)** and RBD **(E)**.

**Figure 3. Structure and function of the spike protein (S protein). (A)** SARS-CoV-2 S protein specifically recognises the ACE2 receptor of the host cells, and thereby starts the infection cycle. **(B)** The S protein undergoes 2 maturation steps by proteolytic cleavage (respectively catalysed by the furin and the TMPRSS2 proteins), which are required to activate the protein and to unlock the fusion peptide. **(C)** Structure of the viral S protein bound to the host ACE2 receptor. The SARS-CoV-2 S protein structure (beige) was produced by running SWISSMODEL on the SARS-CoV homolog (Protein Data Bank entry 6acc), and aligned on the structure of an RBD domain (orange) interacting with ACE2 (grey) from the PDB model 6m0j. The SARS-CoV-2 insertions are highlighted in colors, with a coloring scale reflecting the taxonomic scope of the insertion: red (only found in Human SARS-CoV-2-, yellow, green, blue, and purple (insertion found in most sarbecoviruses.

**Figure 4. Conservation of ACE2 proteins and interactions with the viral S protein. (A)** Interactions between ACE2 and S and conservation of the key residues (adapted from (Wang et al. 2020; Yan et al. 2020)) in different viral strains and animal species. The key interactions between S and ACE2 residues are denoted by solid lines, and weaker interactions by dotted lines. **(B)** Number of differences between human ACE2 and its ortholog in several animal species for the key residues involved in the. Conservation de la protéine ACE2 et interactions avec la protéine S. (adapted from (Yan et al. 2020)).

**Figure 5. Taxonomic coverage of the insertions observed in SARS-CoV-2 S protein.** Each panel shows multiple alignments of amino-acid sequences around the insertion (left) and the likely occurrence of the evolutionary event on the phylogenetic trees inferred from the amino-acid sequences surrounding the insertions (right). The insertions respectively cover the positions 153-158 **(A)**, 245-251 **(B)**, 445-449, **(C)** and 680-683 **(D)** of SARS-CoV-2 S protein. The schema on the top of the panels indicates the respective positions of the 4 insertions. Except for insertion i3b, the sequences sharing a same insertion appear grouped in the phylogenetic tree, suggesting a distinct origin for each insertion. The deep difference between tree topologies indicate that these regions of insertions result from different evolutionary stories. The values on the bifurcations denote the bootstrap score (on a scale from 0 to 100), which indicate the robustness of the corresponding branching. A weak bootstrap value (<50) means that the corresponding branching has a weak reliability. Note that the weak values are often attached to BtYuRmYN02, which results from the metagenomic assembly of a large number of samples for various sources. Consistently, this metagenome is strongly inconsistent between the different aligned fragments, which questions its biological relevance.

**Figure 6. Matches between S gene and HIV genome.** **(A)** Top-ranking alignment between the S gene and the HIV genome. **(B)** Top-ranking alignment between the randomised query sequence (shuffled nucleotides) and the HIV genome. Note the value of the expect score, which indicates the number of false positives expected by chance. The comparison shows that the alignment between the coding sequence of S protein and the HIV genome is not significant. The alignments were performed on NCBI BLAST server (https://blast.ncbi.nlm.nih.gov/Blast.cgi).

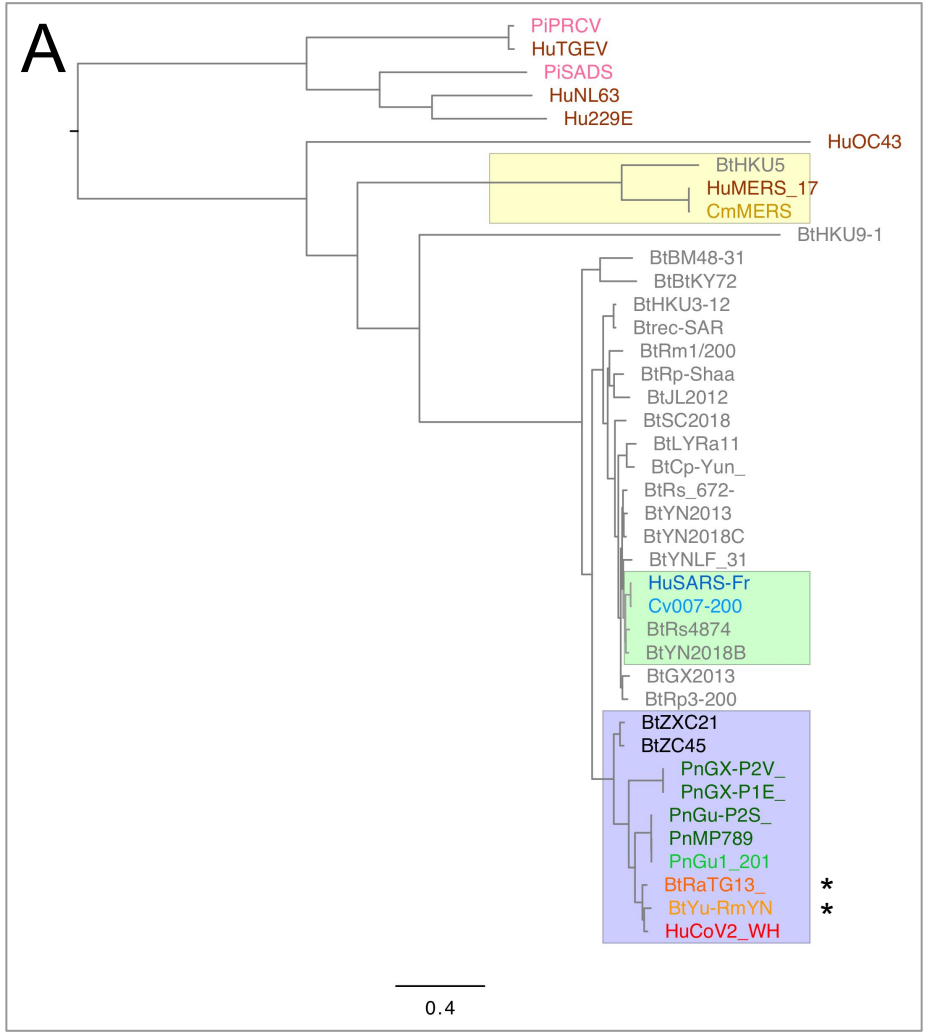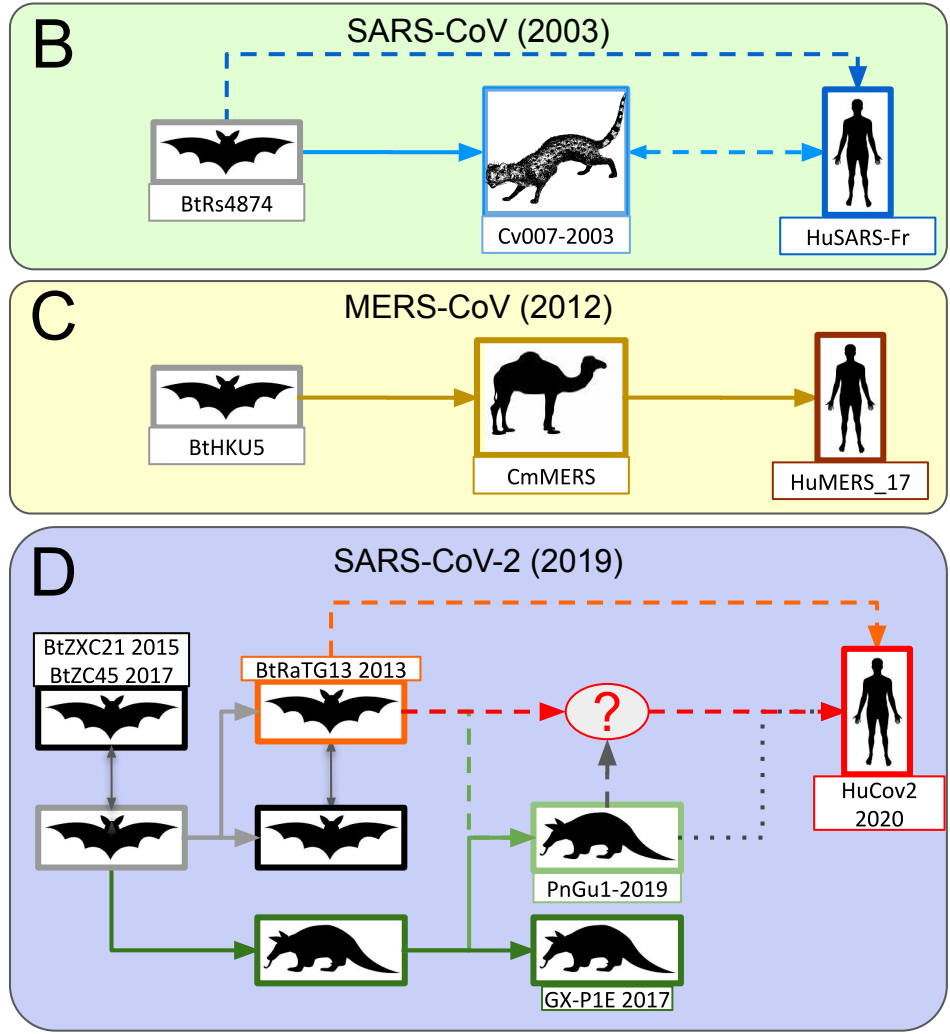

Figure 1

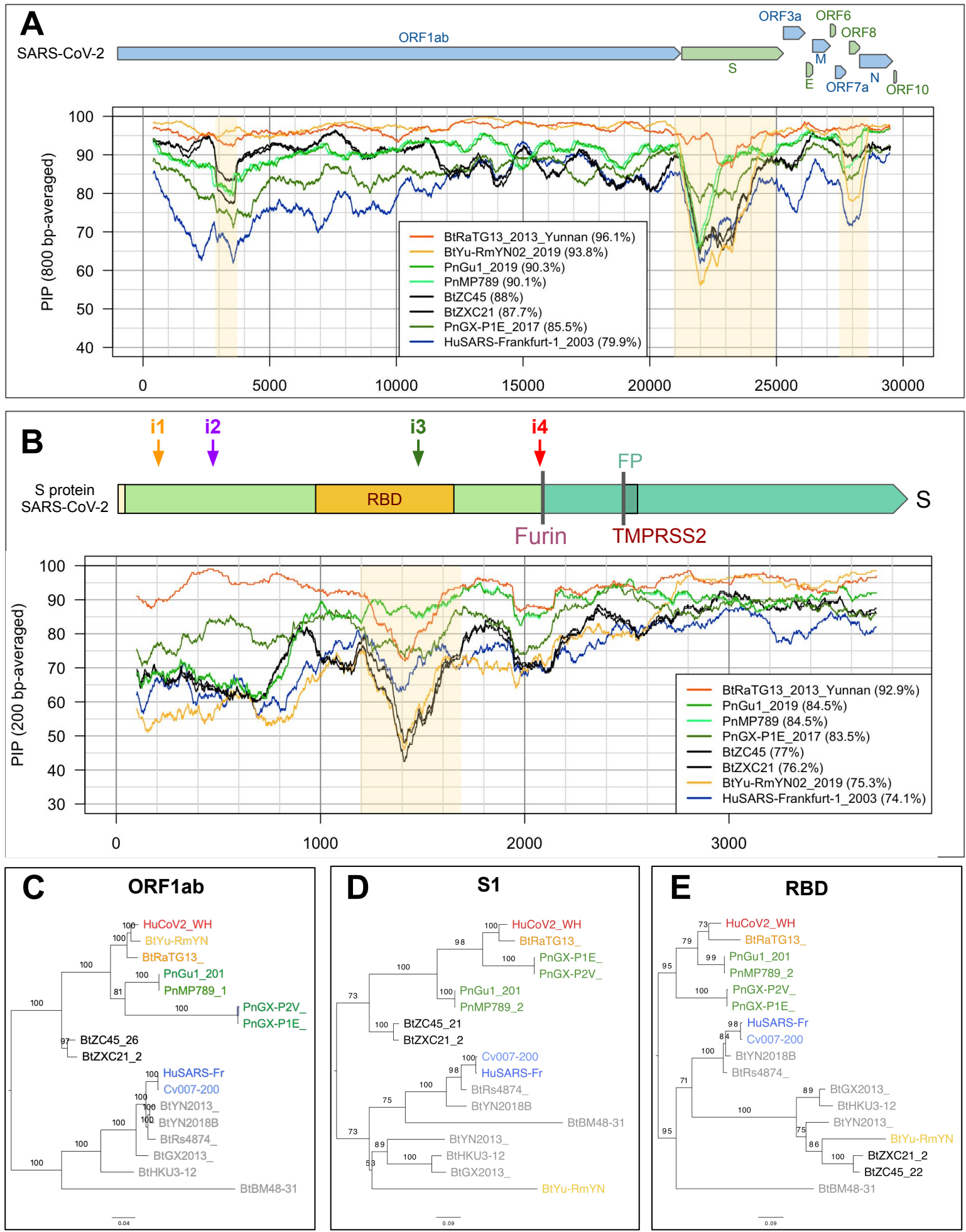

Figure 2

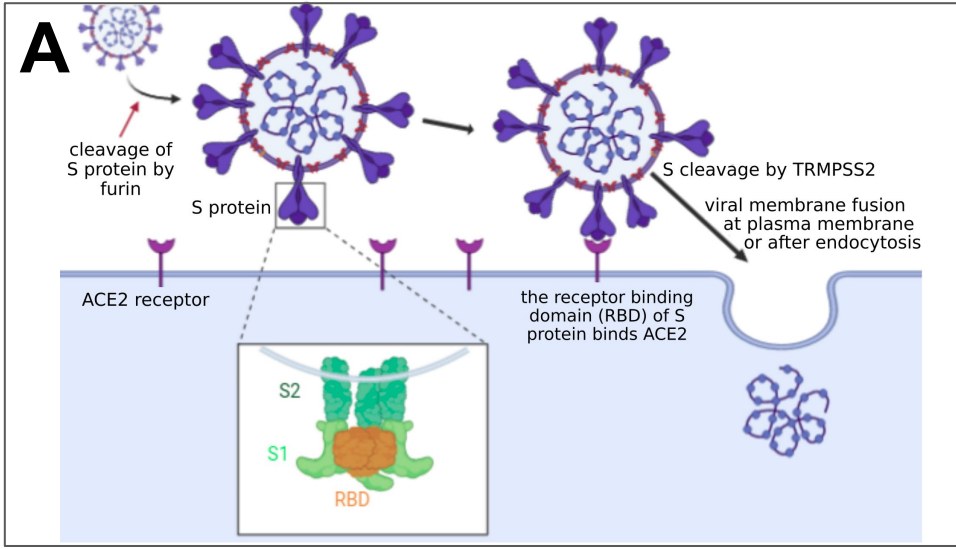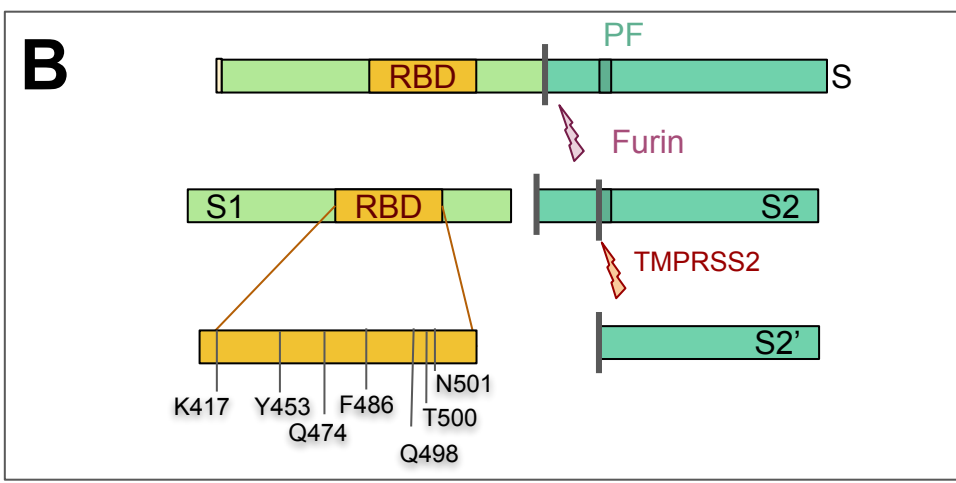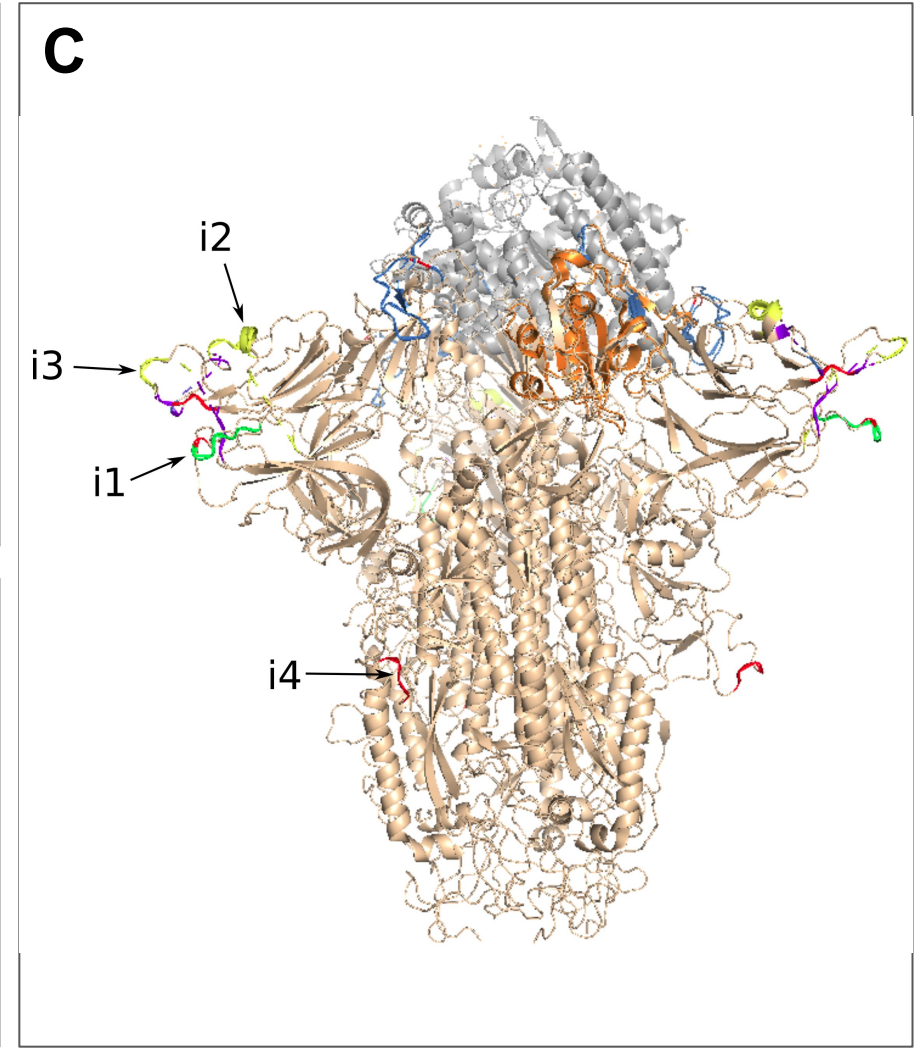

Figure 3

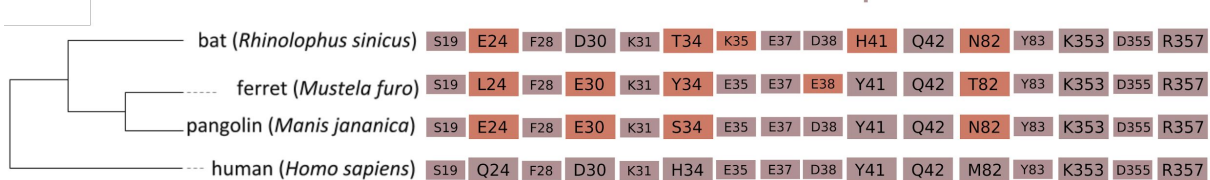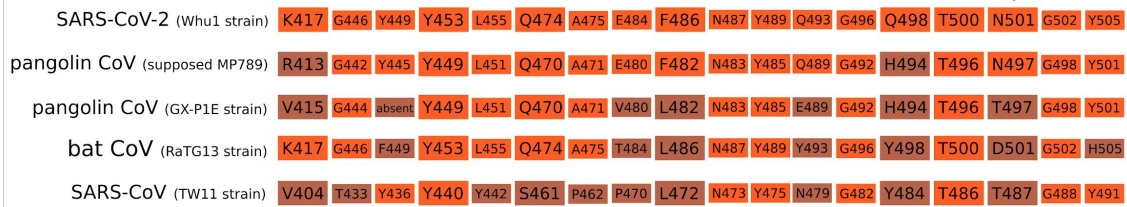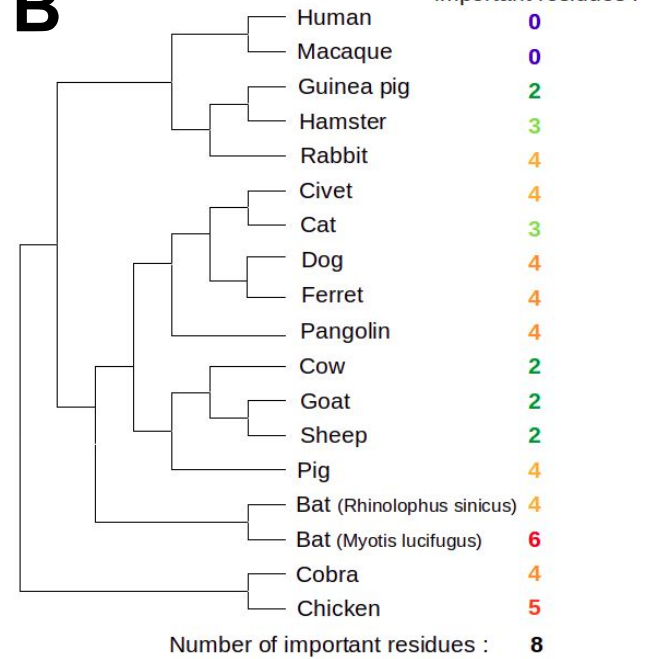

Figure 4

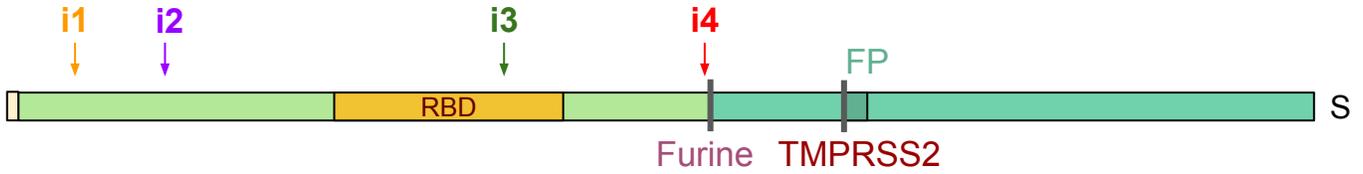
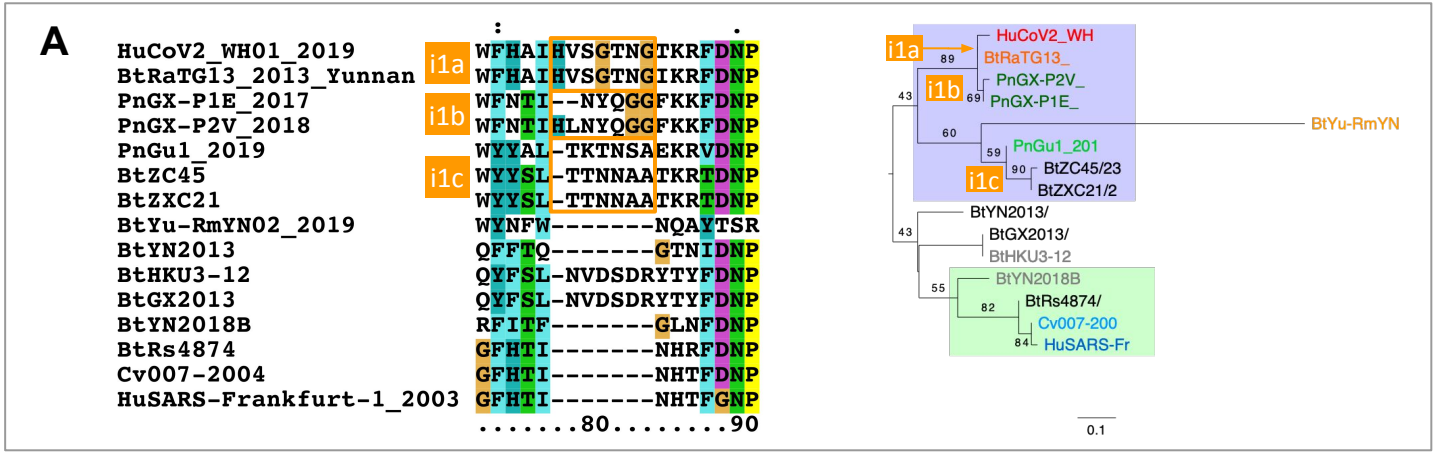
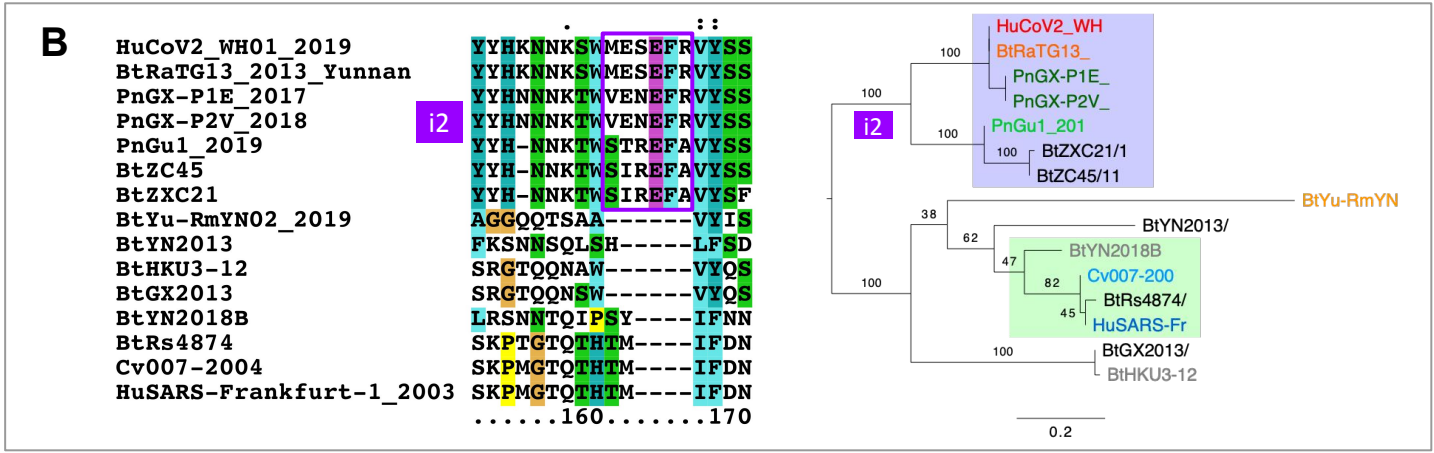
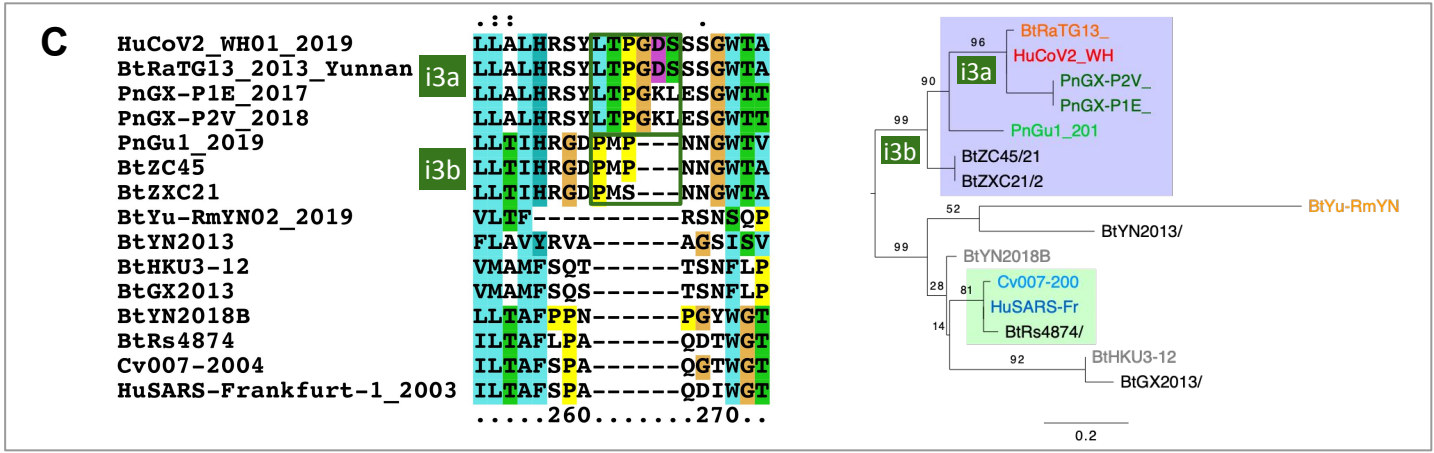
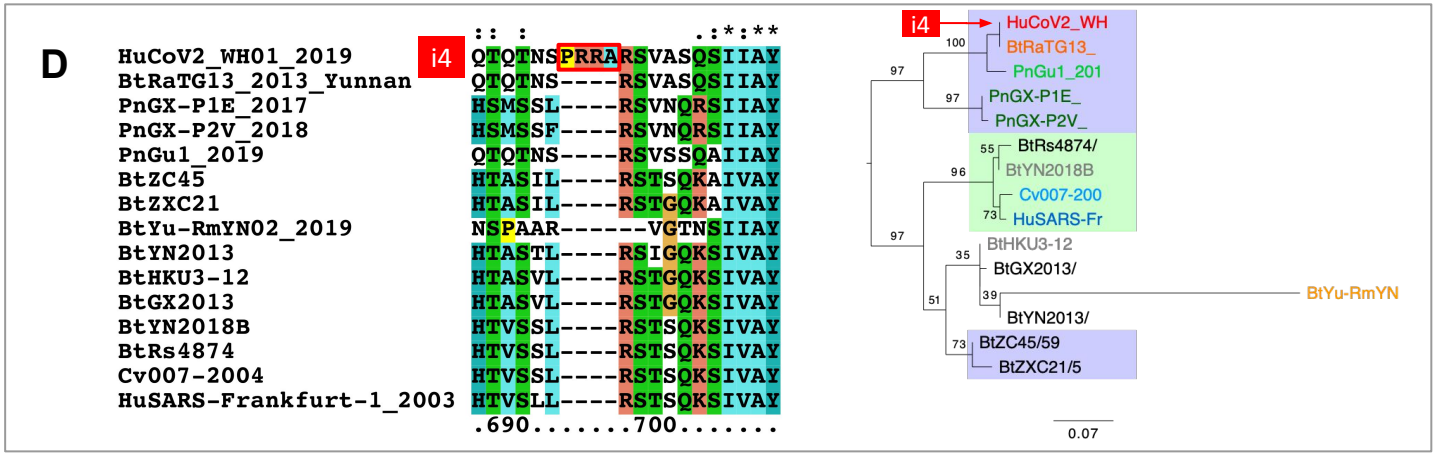

Figure 5

## A

**HIV-1 isolate 19828.PPH11 from Netherlands envelope glycoprotein (env) gene, partial cds**

| Sequence ID: HQ644953.1 | Length: 1143 | Number of Matches: 1 | Range 1: 967 to 994 |
|---|---|---|---|

| Score | Expect | Identities | Gaps | Strand |
|---|---|---|---|---|
| 38.3 bits(41) | 7.5 | 25/28(89%) | 0/28(0%) | Plus/Plus |

```
Query  86    AATGGTACTAAGAGGTTTGATAACCCTG   113
             |||||||||||| ||||| |||||| |||
Sbjct  967   AATGGTACTAAAAGGTTAGATAACACTG   994
```

## B

**HIV-1 isolate patient B clone 16.3 from Netherlands envelope glycoprotein (env) gene, complete cds**

| Sequence ID: HQ386166.1 | Length: 2580 | Number of Matches: 1 | Range 1: 2493 to 2523 |
|---|---|---|---|

| Score | Expect | Identities | Gaps | Strand |
|---|---|---|---|---|
| 39.2 bits(42) | 2.1 | 27/31(87%) | 0/31(0%) | Plus/Minus |

```
Query  351   CCTAAAAGTTCTTTGTAATAACTGTATTATT   381
             ||||||||||||||||||||||  |  ||| |||
Sbjct  2523  CCTAAAAGTTCTTTGTAATATTTCTATAATT   2493
```

Figure 6